# Keynesian models of depression. Supply shocks and the COVID-19 Crisis.

Escañuela Romana, Ignacio[1]

**Abstract.**

The objective of this work is twofold: to expand the depression models proposed by Tobin and analyse a supply shock, such as the Covid-19 pandemic, in this Keynesian conceptual environment. The expansion allows us to propose the evolution of all endogenous macroeconomic variables. The result obtained is relevant due to its theoretical and practical implications. A quantity or Keynesian adjustment to the shock produces a depression through the effect on aggregate demand. This depression worsens in the medium/long-term. It is accompanied by increases in inflation, inflation expectations and the real interest rate. A stimulus tax policy is also recommended, as well as an active monetary policy to reduce real interest rates. On the other hand, the pricing or Marshallian adjustment foresees a more severe and rapid depression in the short-term. There would be a reduction in inflation and inflation expectations, and an increase in the real interest rates. The tax or monetary stimulus measures would only impact inflation. This result makes it possible to clarify and assess the resulting depression, as well as propose policies. Finally, it offers conflicting predictions that allow one of the two models to be falsified.

## 1. Object and results.

This work expands on Tobin's Keynesian models (1975), analyses their local stability, and studies their evolution in the face of a supply shock (specifically, the Covid-19 pandemic).

First, an equation for the real interest rate, based on Taylor's curve, is added. The central bank follows the instrumental objective of the real interest rate. The local stability conditions are studied: the introduction of the interest rate does not make the system more unstable than Tobin's initial system.

Second, the supply shock is modelled as a significant reduction in aggregate full employment output (or natural level of employment). The Covid-19 pandemic is a negative supply shock that represents a sharp drop in the production possibilities of an economy. In this case, the production and/or sale of any service or that requires social proximity, a key factor in the spread of the virus.

The result achieved indicates that the supply shock generates a depression. The force of this depression, as well as the effects on prices and inflation expectations, depend crucially on the current adjustment dynamics. The difference between the predictions would allow the models to be empirically falsified.

---

[1] Doctoral candidate. Universidad de Loyola Andalucía. iescanuelaromana@al.uloyola.es

i. If the adjustment is of the Walras-Keynes-Phillips (WKP) type (Tobin, 1975), via quantities, then a depression is generated that evolves over time: short and long-term effects. Inflation, inflation expectations and the real interest rate rise. It is possible to develop expansionary tax and monetary policies with effects on aggregate demand and, consequently, on the evolution of the output (and therefore, of unemployment or the total number of hours worked in the economy), at the cost of increased inflation.

ii. If the adjustment is of the Marshall (M) type (Tobin, 1975), via prices, then a direct depression occurs, with no differences between short and long-term, reducing inflation and inflation expectations, and increasing the real interest rate. Expansionary tax and monetary policies are totally ineffective in countering the drop in income. The adjustment is only via supply.

## 2. The models.

### 2.1. Conceptual bases.

This work has the following focuses:

I. Macroeconomic. "Macroeconomics … deals with simplified general equilibrium models" (De Vroey, 2004, p.2).
II. Holistic. In the Marshallian sense that the micro-foundation of the model may be left in the background (De Vroey, 2016, p.340).
III. Keynesian[2]. In the sense clarified by Tobin (1993): "it does assert and require that markets not be instantaneously and continuously cleared by prices" (p. 46). This is the central idea: that the economic system may be in a situation of prolonged imbalance with involuntary unemployment. Tobin (1975) develops the possibility that deflation may deepen recessions[3].

Tobin's models consist of a single macroeconomic equilibrium (Tobin, 1993). In the equilibrium, the economy would be in full employment and in the natural rate of unemployment, with constant prices and expectations.

Tobin (1975) starts from the consideration made by Friedman (1971, p. 18): Keynes "assumed that, at least for changes in aggregate demand, quantity was the variable that adjusted rapidly, while price was the variable that adjusted slowly, at least in a downward direction". From here, he proposes two general macro-equilibrium models. The first model is based on the quantity or Keynesian adjustment (the Tobin I model). The second model is the

---

[2] Avoiding embarking on "the troubled path that has led us to forget so much of what Keynes taught us" (Krugman, 2011, p.2).
[3] "The possibility of protracted unemployment which the natural adjustment of a market economy remedy very slowly if at all" (Tobin, 1975, p. 196).

Marshall adjustment (the Tobin II model). Here, the classical dichotomy is valid[4], monetary variables are totally independent from real ones, and conversely.

The Tobin I model: the difference between aggregate demand and aggregate supply impacts the income growth rate. Consequently, insufficient demand causes a drop in output (adjustment of quantities). The difference between aggregate production and its level of full employment has effects on the inflation rate.

The Tobin II model: the difference between the level of aggregate full employment production and the effective output generates the income growth rate. Moreover, the distance between aggregate demand and real production impacts the level of inflation. Insufficient demand causes a drop in prices (adjustment of prices).

In model I, the recessions are the result of insufficient effective demand. In model II, they are the result of an aggregate production that is higher than the potential.

Both models lacked an interest rate equation. However, the interest rate is an endogenous variable that is determined within the system. The model is completed by adding it. To do this, an abbreviated Taylor equation is taken. This results in two models: the expanded Tobin I model (TMIA) and the expanded Tobin II model (TMIIA).

**2.2. Aggregate demand.**

This starts from the definition of aggregate demand (Tobin, 1975).

$$E = C\left(Y_+, Y^*_+, T_-, r_-, \frac{xM}{P}_+, W_+\right) + I(Y_+, Y^*_+, K_-, R_-) + G \tag{1}$$

E is aggregate real effective demand, Y is aggregate real output (income), Y* is its value at full employment, T is taxes minus transfers, r is real interest rate, M is nominal stock of outside money, p or price level, x or expected rate of change of price level, W is private wealth, K or capital stock, G is public spending. Assumed closed economy for simplification.

$$W = \frac{M}{p} + qK \tag{2}$$

Tobin's q is "the ratio of market valuation of capital equity to replacement cost" (op. cit., p. 197). q falls if r rises relative to the marginal efficiency of capital. This efficiency depends positively on Y and Y*, negatively on K.

  a. Price level effect ($E_p$) is negative. On the one hand, Keynes effect: decreasing prices mean larger real money quantity. On the other hand, Pigou effect: higher the real value of private wealth.
  b. $E_x$ is positive, since a rise in inflation expectations decreases the real interest rate (Tobin-Mundell effect)[5].
  c. $E_y$ is greater than 0 but less than 1.

---

[4] Mankiw (1989): the classical dichotomy implies "nominal variables do not affect real variables, the money market is not very important. This classical view of the economy suggests that, for most policy discussions, the money market can be ignored" (p.80).
[5] A reduction in the inflation expectations generates a higher demand for real money balances, which makes the real interest rate rise (Palley, 2005).

d.  $E_{Y^*} < 0$ The fall in natural output makes consumption and investment decrease.

## 2.3. Expansion of the models.

An equation for the real interest rate, based on Taylor's curve (1993), is added. The IS-LM approach traditionally started from the assumption that the central bank proposes the money supply as a target[6]. This proposal can be changed for another whose instrumental target is the real interest rate, which is a more realistic explanation of the effective operation of central banks (Romer, 2000). The real interest rate would be set based on the central bank's output gap targets (deviation of real GDP from a target, Taylor, 1993) and inflation with respect to the price setting target. Therefore, the real interest rate will evolve by:

$$\dot{r} = F((Y - Y^*)_+, (\pi - \pi^o)_+) \tag{3}$$

For simplicity, $\pi^o$ is 0. Moreover, $\pi$ can be changed to x.

Now, aggregate demand depends on four endogenous variables: Y, p, x, r.

## 2.4. TMIA.

Four dynamic model equations:

$$\dot{Y} = A(E - Y) \tag{4}$$
$$\pi = B(Y - Y^*) + x \tag{5}$$
$$\dot{x} = C(\pi - x) = BC(Y - Y^*) \tag{6}$$
$$\dot{r} = D_1(Y - Y^*) + D_2(\pi) = (D_1 + D_2 B)(Y - Y^*) + D_2 x \tag{7}$$

The first three dynamic equations come from Tobin (1975). The second equation can be written as:

$$\dot{p} = pB(Y - Y^*) + px \tag{8}$$

A, B, C, $D_1$ and $D_2$ are parameters.

The ideas in the model are as follows:

i. Production increases due to the difference between the planned demand and the current output. The key idea is Keynesian: that short-term output responds to variations in demand (Tobin 1975, p. 198).
ii. Nominal prices follow expectations plus or minus a "Phillips curve" adjustment to the difference between actual and full employment output (Tobin, 1993).

---

[6] Tobin outlines that r depends on M/p, x, Y, W.

    iii.    Price change expectations are adaptive.
    iv.    The fourth equation is a simplified Taylor curve. Changing $\pi$ for x only changes the quantification of the coefficient.

## 2.5. TMIIA.

The TMIIA equations:

$$\dot{Y} = A'(Y^* - Y) \tag{9}$$

$$\dot{p} = pB'(E - Y) + px \tag{10}$$

$$\dot{x} = C'(\pi - x) = B'C'(E - Y) \tag{11}$$

$$\dot{R} = D_1'(Y - Y^*) + D_2'(x) \tag{12}$$

The ideas in the model are:

    i.    Production increases due to the difference between full employment development and production. Demand does not impact growth.
    ii.    Nominal prices follow expectations plus or minus an adjustment to the difference between demand and output (Tobin, 1993).
    iii.    Price change expectations are adaptive.
    iv.    The fourth equation is a simplified Taylor curve. For simplification, x is considered.

## 2.6. Local stability analysis of the expanded models.

Equilibrium conditions:

$$E - Y = 0 \tag{13}$$

$$Y - Y^* = 0 \tag{14}$$

$$\pi - \pi^* = 0 \tag{15}$$

$$x = 0 \tag{16}$$

$$r - r^* = 0 \tag{17}$$

First, linear approximation of the TMIA system around equilibrium:

$$\begin{pmatrix} \dot{Y} \\ \dot{p} \\ \dot{x} \\ \dot{r} \end{pmatrix} = \begin{pmatrix} A(E_y - 1) & AE_p & AE_x & AE_r \\ pB & 0 & p* & 0 \\ BC & 0 & 0 & 0 \\ D_1 + D_2B & 0 & D_2 & 0 \end{pmatrix} \begin{pmatrix} Y - Y^* \\ p - p^* \\ x - x^* \\ r - r^* \end{pmatrix} \tag{18}$$

The significant necessary condition:

$$-pBE_p - (D_1 + D_2B)E_r > BCE_x \tag{19}$$

This necessary condition is one of reasonable compliance. Only if the Tobin-Mundell effect was relatively strong, and the price level and the interest rate effects were relatively weak, the system would be unstable. The introduction of the interest rate makes the system more stable than Tobin's initial system[7].

Second, linear approximation of the TMIIA system around equilibrium (equations 13 to 17):

$$\begin{pmatrix}\dot{Y}\\\dot{p}\\\dot{x}\\\dot{r}\end{pmatrix} = \begin{pmatrix} -A' & 0 & 0 & 0 \\ pB'E_y - pB' & pB'E_p & pB'E_x + p & pB'E_r \\ B'C'E_y - CB & C'B'E_p & C'B'E_x & C'B'E_r \\ D'_1 & 0 & D'_2 & 0 \end{pmatrix}\begin{pmatrix}Y-Y^*\\p-p^*\\x-x^*\\r-r^*\end{pmatrix} \tag{20}$$

Necessary conditions relevant to local stability:

$$-pE_p > C'E_x \tag{21}$$

This is the Tobin's necessary condition. This is met if the positive effect of inflation on demand (Keynes and Pigou effects) is stronger than the Tobin-Mundell effect.

## 3. A supply shock. The pandemic.

A supply shock increases or reduces the aggregate productive capacity or full employment production. For example, if rainfall increases in a territory, this leads to an increase in the production possibilities for certain agricultural productions and associated services.

The Covid-19 pandemic operates as an exogenous negative supply shock with a highly significant impact, which reduces the potential production of goods and services dependent on social proximity. These are affected, especially the longer it goes on.

This makes a part of the installed capacities and trained human capital redundant. They could produce goods and services, but the risk of contagion prevents them.

The reduction in potential production makes effective production operate beyond its possibilities. The economy is in a situation of overheating that is compatible with the existence of a very significant unemployment: $\dot{Y}^* < 0$ (Y-Y* increases, Y*-Y decreases). The economy is operating above its threshold of possibilities in the short-term, not being able to produce or generate goods and services that are impossible to sell. The natural rate of unemployment rises.

---

[7] $pE_p + CE_x < 0$

The models must answer three crucial questions: What are the short-term effects? What are the long-term effects? Which policies should be used to combat the supply shock?

## 5. Prediction of the models in the face of a supply shock.

### 5.1. TMIA.

In the short term, the reduction in the potential output negatively affects consumption and investment: $E_{Y^*} < 0$ This generates an economic depression if the negative supply shock is relevant: $\dot{Y} < 0$ (equation (4)). Simultaneously, inflation and inflation expectations increase ($\dot{p}, \dot{x} > 0$), since the distance between actual and potential output has increased and the economy is in a situation of overheating (equations (5) and (6)). This fact and the increasing inflation produce an increase in the real interest rate ($\dot{r} > 0$), equation (7).

In the long term, all endogenous variables interrelate with all the others. The increases in prices and interest rates generate a reinforcement in the fall in aggregate real income, since $E_p$ and $E_r$ are negative. Therefore, national income falls more profoundly ($\dot{Y} < 0$) as increases in inflation and the real interest rate confirm the depression. The rise in inflation expectations cushions the fall, but we must remember that in order to offset the other effects, the model would have to be locally unstable. This is also unreasonable. Globally, national income Y decreases, dropping the economy into recession until Y approaches the new Y*.

The movement in the endogenous variables leads to a new equilibrium, provided that the established stability conditions are met.

Keep in mind that the adjustment is intuitively contrary to the usual one. Generally, aggregate demand E would autonomously boost aggregate production Y. E now follows a supply shock.

The result is a new equilibrium: $(Y = Y^*_-, p = p^*_+, x = x^*_+, r = r^*_+)$

It is a ceteris paribus result: in the absence of other exogenous modifications. What effects would expansionary tax and monetary policies have?

i. An expansionary public spending policy would boost demand E and, consequently, the growth of income Y ($E_G > 0$). The effects would be a cushioning in the fall of Y, sustaining the distance with respect to the full employment output. There would also be a further increase in inflation, expectations, and real interest rates.
ii. A relaxed monetary policy would determine a reduction in real interest rates. This again boosts demand E through consumption and investment ($E_r < 0$).
iii. A positive result from sustaining demand with both policies, in the face of a supply shock like the pandemic, is that production is sustained and unemployment is reduced, gaining time until a vaccine is found or waiting until the production and consumption processes are adapted to be compatible with greater social distancing. The cost is increased inflation (and public debt) since $Y > Y^*$.

## 5.2. TMIIA.

The aggregate output is reduced ($\dot{Y} < 0$) in direct proportion to the drop in potential production. The fall depends on parameter $A'$ (equation (9)). The depression is a supply-side event and the result of a constant negative slope. Inflation and inflation expectations decrease ($\dot{p}, \dot{x} < 0$) because of the falling aggregate demand $E_{Y^*} < 0$ (equations (10) and (11)). The real interest rates grow ($\dot{r} > 0$), because of the growing distance between the actual output and the equilibrium value (equation (12))[8]. The following impacts of inflation, expectations and interest rates are reduced to the evolution of prices and their expectations. The depression does not move with the nominal variables.

Therefore, all movements in prices and interest rates are irrelevant to production, whose fall is linear. We are seeing a classic adjustment: only production moves and solely due to supply factors. The adjustment and its speed are independent from demand. Finally, expansionary fiscal and monetary policies would not have any impact on output.

The other three endogenous variables are adjusted through the impacts on aggregate demand. The new equilibrium will be: $(Y = Y^*_-, p = p^*_- , x = x^*_-, r = r^*_+)$.

## 6. Conclusions.

Firstly, Tobin's depression models have been expanded, adding an equation for the interest rate to include all endogenous variables. The conditions for local stability have been analysed. The four-equation models are not more unstable than Tobin's initial system.

Second, the problem of the supply shock has been raised in these models, to analyse what macroeconomic effects they predict. A supply shock increases or reduces the aggregate productive capacity, changing the natural level of unemployment. The Covid-19 pandemic operates as a strong exogenous negative supply shock which reduces the potential production of goods and services dependent on social proximity.

Third, the TMIA (the quantity or Keynesian adjustment) and TMIIA (The Marshallian adjustment via prices) models give different predictions:

   a) The channels through which the supply shock is transmitted (demand versus supply) are different.
   b) The speed of the economic adjustment to the shock is dissimilar. TMIA predicts differences between the short and long-term. TMIIA does not.
   c) Inflation and inflation expectations increase in TMIA and decrease in TMIIA. This allows the models to be empirically falsified.
   d) An expansionary tax policy and a relaxed monetary policy are positives for TMIA and ineffective in TMIIA. TMIIA contains the traditional doctrine that money is a veil with no real effects (classical dichotomy).

Fourth, the TMIA predicts an initial depression, in the short-term, whose amount depends on the value of the derivative of demand with respect to output ($E_y$). The greater this derivative,

---

[8] Although the fall in price expectations cushions this increase.

the more income spent on consumption and investment, the lower the impact in the short-term.

Moreover, inflation and real interest rates increase. In the medium and long-term, these increases will worsen the depression. Given the growth in inflation expectations, the Tobin-Mundell effect will partially cushion the situation, preventing a certain reduction in the real interest rate. We must remember that if this latter effect were very intense, the economy would not tend to equilibrium.

In the case of the Covid-19 pandemic, expansionary fiscal and monetary policies can sustain demand and maintain low real interest rates. Their impact is cushioning, and they gain time. However, the cost is higher inflation since output is above its potential level. Therefore, these demand measures must be prevented from masking the need for the political supply measures. Overcoming the disease (vaccine) and/or restructuring measures on the supply side so that social distancing does not prevent the provision of services and the manufacturing of goods. Some redistribution of physical and human resources between sectors would be necessary.

Fifth, the TMIIA predicts a drop in linear output. There is no difference between the short and long-term, as income linearly varies depending on the difference between full employment output and effective output. All demand policies that are taken lead solely to a price increase. Expansionary policies impact on the nominal variables, but not on the real ones.

In the case of the Covid-19 pandemic, this means that the recovery of income can only come from supply factors: either from overcoming the disease, or from applying restructuring measures.

I leave for further research the following questions:

1. TMIA. The conditions under which inflation could rapidly increase. This possibility could be produced by expansionary policies. Among others, the impact on the real money supply and on the real private wealth could be so strong that it ended up offsetting the positive impacts of the low real interest rate.
2. The expansion of these models to open economies.
3. An econometric analysis. When there is a negative supply shock, do inflation and expectations increase or decrease?

**References.**